\documentclass[12pt]{article}
\usepackage{epsfig}

\usepackage{a4}

\usepackage{amsmath}

\newcommand{\insertplot}[5]{\begin{figure}
 \hfill\hbox to 0.05in{\vbox to #5in{\vfill
 \inputplot{#1}{#4}{#5}}\hfill}
 \hfill\vspace{-.1in}
 \caption{#2}\label{#3}
 \end{figure}}
 \newcommand{\inputplot}[3]{
 \special{ps: plotfile #1}
}
\newcounter{fig}   

\usepackage{epsfig}
\usepackage{amsmath}
\usepackage{amsfonts}
\usepackage{graphicx}
\usepackage[german, english]{babel}
\usepackage{a4wide}
\usepackage{amsmath}
\usepackage{amssymb}
\usepackage{ifthen}
\usepackage{epsfig}

\begin{document}

\newcommand{\dd}{\mbox{d}}
\newcommand{\tr}{\mbox{tr}}
\newcommand{\la}{\lambda}
\newcommand{\ta}{\theta}
\newcommand{\f}{\phi}
\newcommand{\vf}{\varphi}
\newcommand{\ka}{\kappa}
\newcommand{\al}{\alpha}
\newcommand{\ga}{\gamma}
\newcommand{\de}{\delta}
\newcommand{\si}{\sigma}
\newcommand{\bomega}{\mbox{\boldmath $\omega$}}
\newcommand{\bnabla}{\mbox{\boldmath $\nabla$}}
\newcommand{\bsi}{\mbox{\boldmath $\sigma$}}
\newcommand{\bchi}{\mbox{\boldmath $\chi$}}
\newcommand{\bal}{\mbox{\boldmath $\alpha$}}
\newcommand{\bpsi}{\mbox{\boldmath $\psi$}}
\newcommand{\brho}{\mbox{\boldmath $\varrho$}}
\newcommand{\beps}{\mbox{\boldmath $\varepsilon$}}
\newcommand{\bxi}{\mbox{\boldmath $\xi$}}
\newcommand{\bbeta}{\mbox{\boldmath $\beta$}}
\newcommand{\be}{\begin{equation}}
\newcommand{\bea}{\begin{eqnarray}}
\newcommand{\ii}{\mbox{i}}
\newcommand{\e}{\mbox{e}}
\newcommand{\pa}{\partial}
\newcommand{\Om}{\Omega}
\newcommand{\vep}{\varepsilon}
\newcommand{\bfph}{{\bf \phi}}
\newcommand{\lm}{\lambda}
\def\theequation{\arabic{equation}}
\renewcommand{\thefootnote}{\fnsymbol{footnote}}
\newcommand{\re}[1]{(\ref{#1})}
\newcommand{\R}{{\rm I \hspace{-0.52ex} R}}
\newcommand{\N}{{\sf N\hspace*{-1.0ex}\rule{0.15ex}%
{1.3ex}\hspace*{1.0ex}}}
\newcommand{\Q}{{\sf Q\hspace*{-1.1ex}\rule{0.15ex}%
{1.5ex}\hspace*{1.1ex}}}
\newcommand{\C}{{\sf C\hspace*{-0.9ex}\rule{0.15ex}%
{1.3ex}\hspace*{0.9ex}}}
\newcommand{\eins}{1\hspace{-0.56ex}{\rm I}}
\renewcommand{\thefootnote}{\arabic{footnote}}
\newcommand{\ee}{\end{equation}}
\newcommand{\eea}{\end{eqnarray}}

\title{ Scalar hairy black holes and solitons \\in a gravitating Goldstone model}
\author{
{\large Eugen Radu}$^{\dagger \star}$,
{\large Ya. Shnir}$^{  \ddagger}$
 and {\large D. H. Tchrakian}$^{\dagger \star}$ \\ \\
$^{\dagger}${\small  School of Theoretical Physics -- DIAS, 10
Burlington Road, Dublin 4, Ireland}
\\ $^{\star}${\small Department of Computer Science, National
University of Ireland Maynooth, Ireland}
\\ $^{\ddagger}${\small Department of Mathematical Sciences, Durham University, UK}}

\maketitle

\begin{abstract}
We study  black hole solutions of Einstein gravity coupled to a specific global symmetry breaking Goldstone model described by an
$O(3)$ isovector scalar field in four spacetime dimensions. Our configurations are static and spherically symmetric,  approaching at
infinity a Minkowski spacetime background. A set of globally regular, particle-like solutions
are found in the limit of vanishing event horizon radius. These configurations can be viewed as {\it 'regularised'}
global monopoles, since their mass is finite and the spacetime geometry has no deficit angle.
As an unusual feature, we notice the existence of extremal black holes in
this model defined in terms of gravity and scalar fields only.
\end{abstract}

%\medskip

%%%%%%%%%%%%%%%%%%%%%%%%%%%%%%%%%%%%%%%%%%%%%%%%%%%%%%%%%%%%%%%%%%
\section{Introduction}
%%%%%%%%%%%%%%%%%%%%%%%%%%%%%%%%%%%%%%%%%%%%%%%%%%%%%%%%%%%%%%%

Black holes with scalar hair are rather a common presence in the
lanscape of gravity solutions with anti-de Sitter asymptotics.
Some of these configurations have found interesting applications
in the context of gravity/gauge duality, see $e.g.$ \cite{Gubser:2008px}.
The situation is, however, rather different in the absence of
a cosmological constant. Asymptotically flat black holes in models featuring scalar fields
are rather scarce. Such solutions have been mainly studied as counterexamples to
the no hair conjecture \cite{RW} and typically contain also gauge fields (for a review of this topic,
see \cite{Volkov:1998cc}). Interestingly, as proven in the case of Einstein-Skyrme theory,
there are hairy black hole solutions even in theories with scalar fields only \cite{gskyrm}.
These Einstein-Skyrme solutions were shown to be stable \cite{Heusler:1992av}, \cite{Maeda:1993ap}.

Perhaps the simplest examples of black hole solutions in a model with scalar fields only
are found in a symmetry breaking model featuring an $O(3)$ scalar isovector field.
These are the black holes inside the global monopoles \cite{Barriola:1989hx},
which were discussed  in \cite{Liebling:1999ke}, \cite{Maison:1999ke}.
Global monopoles are topological defects that arise in certain theories where a global symmetry is
spontaneously broken. Like the well-known 't Hooft-Polyakov monopoles, these configurations
are constructed within an hedgehog-type Higgs field Ansatz
and possess a unit conserved topological charge, which is the winding number of the scalars.
% but no scalar charge as defined from the far field asymptotics.
However, for both solutions with a regular origin
and black holes, the energy density decays as $1/r^2$ at large distances
and hence their masses, defined in the usual way, diverge.
This leads to a deficit solid angle in the geometry of the space
and the resulting spacetime is not strictly asymptotically flat\footnote{See Ref. \cite{Nucamendi:1996ac} for
a discussion of this type of asymptotics together with a definition of the mass through the application
of the standard Hamiltonian analysis. Global monopoles and black holes in Einstein-Goldstone model with a cosmological constant
are studied $e.g.$ in \cite{Brihaye:2005qr}. The mass and action of the solutions are also computed there by using
a boundary counterterm subtraction method.}.

Hairy black holes with finite mass approaching  at infinity a Minkowski spacetime background
are found in the gauged generalisation of this model, with a non-Abelian (local) gauge group $SO(3)$.
This, of course, is the usual Georgi-Glashow model supporting 't Hooft-Polyakov monopoles.
These configurations were extensively studied in the literature, from various directions \cite{gmono}.
In contrast to the ungauged version, the black holes inside 't Hooft-Polyakov monopoles
become extremal in a critical limit.
Heuristically, this property can be attributed to the existence of a magnetic charge in this model.
As is well known, this charge is completely specified
in terms of the scalar triplet of Higgs fields \cite{Arafune:1974uy}.

This leads to the interesting question, as to whether one can find finite mass solutions with similar properties
in a simple model with a scalar isovector field only, $i.e.$ without gauge fields.
Such solutions would still possess a 'magnetic'-type topological charge and thus may allow for extremality;
however, their existence would require a deviation from the standard scalar fields action.
In this work we answer this question by constructing solutions of a specific Goldstone
model in 3+1 dimensions, originally proposed in \cite{Tchrakian:1990ai}.
Our solutions have finite mass and approach a Minkowski spacetime background at infinity.
They also share a number of basic properties with the gravitating 't Hooft-Polyakov monopoles.
In particular, and in strong contrast with the usual global monopoles with an event horizon in
\cite{Liebling:1999ke}, \cite{Maison:1999ke}, we find that extremal black holes exist even in a
model with  scalar fields only.

%%%%%%%%%%%%%%%%%%%

%%%%%%%%%%%%%%%%%%%%%%%%%%%%%%%%%%%%%%%%%%%%%%%%%%%%%%%%%%%%%%%
\section{The model}
%%%%%%%%%%%%%%%%%%%%%%%%%%%%%%%%%%%%%%%%%%%%%%%%%%%%%%%%%%%%%%%

%%%%%%%%%%%%%%%%%%%%%%%%%%%%%%%%%%%%%%%%%%%%%%%%%%%%%%%%%%%%%%%%%%
 \subsection{Action and field equations}
%%%%%%%%%%%%%%%%%%%%%%%%%%%%%%%%%%%%%%%%%%%%%%%%%%%%%%%%%%%%%%%

We consider the 3+1 dimensional action
\be
\label{act}
S=\int d^4x \sqrt{-g} \left(\frac{R}{16 \pi G} - L_{m} \right),
\ee
where the gravity part of the action is the usual Einstein-Hilbert action with curvature
scalar $R$ and   $G$
  the gravitational coupling constant.
  $L_{m}$ is
the Lagrangian of the matter fields, which is given by a
symmetry breaking model in 3+1 spacetime dimension to which we refer
as a Goldstone model. In general, $L_{m}$ contains three different parts,
\begin{eqnarray}
\label{L-gold}
 L_{m}=\lambda_1 U(|\vec \phi|)  \partial _{\mu}\vec \phi \cdot \partial^{\mu}\vec \phi
 +\lambda_2 (\partial _{\mu}\vec \phi \times \partial_{\nu}\vec \phi)^2
 +V(|\vec \phi|),
\end{eqnarray}
%where we use the shorthand notation
%\[
%U(|\vec \phi|)=(\eta^2-|\vec \phi|^2)\,,
%\]
 with $\vec \phi\equiv \f^a=(\phi^1,\phi^2,\phi^3)$   a triplet of real scalar fields.
 The first part  above is the usual kinetic term multiplied with a correction factor $U$ which depends
 only on the magnitude of $\vec \phi$; the second part
$(\partial _{\mu}\vec \phi \times \partial_{\nu}\vec \phi)^2$ is
a Skyrme-like term, while $V(|\vec \phi|)$ is
 a symmetry breaking potential.

 The corresponding  Einstein   equations are found by varying
(\ref{act}) with respect to $g_{\mu \nu}$
 and read
 \begin{eqnarray}
\label{Eeq}
R_{\mu \nu}-\frac{1}{2} R g_{\mu\nu}=8 \pi G~ T_{\mu \nu},
 \end{eqnarray}
 with the energy-momentum tensor
 \begin{eqnarray}
 \label{stress-mu-nu}
T_{\mu\nu}&=&
2\,\la_1\,U(|\vec\f|)
\left[\left(\pa_{\mu}\f^a\pa_{\nu}\f^a\right)
-\frac12\,g_{\mu\nu}\,\left(\pa_{\tau}\f^a\pa_{\la}\f^a\right)\,g^{\tau\la}\right]
 -g_{\mu\nu}\,V(|\vec\f|)
\\
&&
+4\la_2\left[\left(\pa_{[\mu}\f^a\pa_{\tau]}\f^b\right)\,\left(\pa_{[\nu}\f^a\pa_{\la]}\f^b\right)\,g^{\tau\la}
-\frac14\,g_{\mu\nu}\,\left(\pa_{[\rho}\f^a\pa_{\tau]}\f^b\right)\,
\left(\pa_{[\si}\f^a\pa_{\la]}\f^b\right)\,g^{\rho\si}\,g^{\tau\la}\right]
\nonumber
\end{eqnarray}
The equation of motion  for the scalar fields is
 \begin{eqnarray}
\label{Geq}
% \la_1U(|\vec\f|)\left[U(|\vec\f|)\pa_{\mu}\pa^{\mu}\f^a-2(\pa_{\mu}|\vec\f|^2)\,\pa^{\mu}\f^a+
%(\pa_{\mu}\vec\f\cdot\pa^{\mu}\vec\f)\,\f^a\right]
%\nonumber
%\\
% \qquad\qquad\qquad\qquad-4\la_2\,\pa^{\nu}\,(\pa_{[\mu}\f^a\,\pa_{\nu]}\f^b)+\la_0\,\pa_{\f_a}V=0\,.
 && \lambda_1\left [
\frac{2}{\sqrt{-g}}\,\partial_{\mu}\left (\sqrt{-g}\,U(|\vec\f|)  \pa^{\mu} \f^a \right)
- (\pa_{\mu}\f^b \pa^{\mu}\f^b) \pa_{\f^a}U(|\vec\f|)
 \right ]
 \\
 \nonumber
&&
 \qquad\qquad
+\frac{4\la_2}{\sqrt{-g}}\,\pa^{\nu}\,\left (\sqrt{-g}\, \pa_{[\mu}\f^a\,\pa_{\nu]}\f^b \right)
- \,\pa_{\f^a}V(|\vec\f|)=0~.
%(\pa_{\mu}\vec\f\cdot\pa^{\mu}\vec\f)\,\f^a\right]
%\nonumber
%\\
% \qquad\qquad\qquad\qquad-4\la_2\,\pa^{\nu}\,(\pa_{[\mu}\f^a\,\pa_{\nu]}\f^b)+\la_0\,\pa_{\f_a}V=0\,.
\end{eqnarray}
As a general feature, the scalar field $\vec \phi$
 is a relic of a Higgs field and
has the same dimensions $L^{-1}$ as a gauge connection.
Asymptotically, is satisfies the
symmetry breaking boundary condition
  \begin{eqnarray}
\label{symm}
\lim_{r\to \infty}|\vec \phi|=\eta.
\end{eqnarray}

%%%%%%%%%%%%%%%%%%%%%%%%%%%%%%%%%%%%%%%%%%%%%%%%%%%%%%%%%%%%%%%
\subsection{The spherically symmetric ansatz }
%%%%%%%%%%%%%%%%%%%%%%%%%%%%%%%%%%%%%%%%%%%%%%%%%%%%%%%%%%%%%%%

In this paper we shall consider spherically symmetric globally regular and
black hole solutions to the system \re{act}.
A generally enough metric ansatz reads
%For the spherically symmetric field configuration it is convenient to use the metric
\begin{eqnarray}
\label{metr-sph}
%ds^2 = -\sigma(r)^2 N(r) dt^2 + N(r)^{-1} dr^2 + r^2(d\theta^2 + \sin^2 \theta d\varphi^2) \,
ds^2 = -f_0(r) dt^2 + f_1(r) dr^2 + f_2(r)(d\theta^2 + \sin^2 \theta d\varphi^2) \, ,
\end{eqnarray}
where $t$ is the time coordinate,
$r$ is the radial coordinate (with $r^2=x^a x^a$),   while $\theta$ and $\varphi$ are the angular coordinates within the
usual range.
In the numerical construction of asymptotically flat configurations,
we have mainly employed the usual
Schwarschild coordinates with
\begin{eqnarray}
\label{metr-sph2}
 f_0(r)=N(r)\sigma(r)^2,~~f_1(r)=\frac{1}{N(r)},~~f_2(r)=r^2,
 ~~~{\rm and}~~N(r)=1-\frac{2m(r)}{r},
\end{eqnarray}
where $m(r)$ may be interpreted as the total
mass-energy within the radius $r$.
For black hole solutions, the event horizon is
at $r = r_h$ where $N(r_h) = 0$ and $\sigma(r_h)>0$.
For solitons, $r=0$ is a regular origin, with $N(0)=1$, $\sigma(0)>0$.

For the scalar field, we use the usual hedgehog ansatz, with
\be
\label{sph}
\f^a=\eta\,h(r)\,\hat x^a\,,
\ee
and $\hat x^a= x^a/r=(\sin \theta \cos \varphi,\sin \theta \sin \varphi,\cos \theta)$.

With this ansatz, the field
equations (\ref{Eeq}), (\ref{Geq}) take the relatively simple form
 (where the prime denotes derivative with respect to $r$):
\begin{eqnarray}
\label{eqs1}
m'=\alpha^2
\left(
Nh'^2 T_1+T_2
\right),
~~\frac{\sigma'}{ \sigma}= \frac{2 \alpha^2}{r} h'^2T_1,
 \end{eqnarray}
for the metric functions, and
 \begin{eqnarray}
\label{eqs2}
(\sigma Nh'T_1)'=\frac{1}{2}\sigma
\left(
Nh'^2 \frac{\partial T_1}{\partial h}+\frac{\partial T_2}{\partial h}
\right),
 \end{eqnarray}
 for the scalar amplitude.
In these relations we define as usual
 \begin{eqnarray}
\label{alpha}
\alpha^2= 4 \pi G \eta^2,
 \end{eqnarray}
and we use the shorthand notation
 \begin{eqnarray}
\label{Ti}
 T_1=\lambda_1 U(h) r^2+2 \lambda_2 h^2,~~T_2=2 \lambda_1 U(h) h^2+\lambda_2 \frac{h^4}{r^2}+r^2 V(h).
 \end{eqnarray}
%and, accordingly,
%  \begin{eqnarray}
%\label{dTi}
%\frac{\partial T_1}{\partial h}=\lambda_1\frac{\partial U}{\partial h}+4 \lambda_2 h,~~
%\frac{\partial T_2}{\partial h}=
%2\lambda_1(h^2\frac{\partial U}{\partial h}+2U h)+4 \lambda_2 \frac{h^3}{r^2}+r^2\frac{\partial V}{\partial h}
% \end{eqnarray}

%%%%%%%%%%%%%%%%%%%%%%%%%%%%%%%%%%%%%%%%%%%%%%%%%%%%%%%%%%%%%%%
%\subsection{The global charge}
%%%%%%%%%%%%%%%%%%%%%%%%%%%%%%%%%%%%%%%%%%%%%%%%%%%%%%%%%%%%%%%
As originally discussed in \cite{Arafune:1974uy},
 the possibility of a nonvanishing `magnetic` charge in
 a model with an $O(3)$ isovector scalar field
 is determined by the homotopy class of the scalar fields only,
 being independent of the gauge fields.
 Moreover, this is true no matter what action principle determines
 the dynamics of $\phi^a$.
 Following  \cite{Arafune:1974uy}, one can define a 't~Hooft `electromagnetic` tensor
\begin{eqnarray}
F_{\mu \nu}= -\epsilon_{abc}\hat \phi^a \partial_\mu \hat \phi^b \partial_\nu \hat \phi^c,
\end{eqnarray}
where $\hat \phi^a=\phi^a/|\vec\f|$.
A straightforward computation shows that for the  hedgehog ansatz (\ref{sph}),
the only nonvanishing component of this tensor is $F_{\theta \varphi}=\cos \theta$, which after integration over $S^2$
gives a unit 'magnetic' charge for the solutions, as expected.
Note that this is a generic feature independent on the coupling with gravity, or,
on the existence of finite energy solutions of the equations
(\ref{eqs1}), (\ref{eqs2}).

%%%%%%%%%%%%%%%%%%%%%%%%%%%%%%%%%%%%%%%%%%%%%%%%%%%%%%%%%%%%%%%
\subsection{A virial identity}
%%%%%%%%%%%%%%%%%%%%%%%%%%%%%%%%%%%%%%%%%%%%%%%%%%%%%%%%%%%%%%%

By expressing the curvature scalar $R$ in terms of the metric function $m(r)$ and $\sigma(r)$
and dropping a total divergence term,
we obtain the following form of the effective Lagrangean of our static spherically
symmetric system:
\begin{eqnarray}
\label{Leff}
L_{eff}=  \sigma \left (m'-\alpha^2   (Nh'^2 T_1+T_2) \right)~.
 \end{eqnarray}

This form of the reduced Lagrangean allow us to obtain an interesting virial relation.
Following the approach proposed  in \cite{Heusler:1996ft},
 let us  assume the existence of a solution $m(r), \sigma(r), h(r)$ with suitable
boundary conditions at the horizon and at infinity.
% (this approach work for finite mass, regular solutions only).
Then each member of the 1-parameter family $F_\lambda(r)\equiv F(r_h+\lambda (r-r_h))$ (with $F=(m,\sigma,h)$
and $\lambda$ some arbitrary real parameter (which should not be confused with
the constants $\lambda_1,~\lambda_2$))
 assumes the same boundary values at $r = r_h$ and $r = \infty$. Then the action
 $S[m_\lambda,\sigma_\lambda,h_\lambda ]$
 must have a critical point at $\lambda = 1,$ $[dS/d\lambda]|_{\lambda=1} = 0$. The result
 is the following virial relation
\begin{eqnarray}
\label{vir1}
\int_{r_h}^{\infty}dr
(
{\cal P}_0+\lambda_1 {\cal P}_1+\lambda_2 {\cal P}_2
)
=0,
\end{eqnarray}
where
\begin{eqnarray}
\nonumber
&&{\cal P}_0=\sigma  r^2 V(h)(3-\frac{2r_h}{r} ),~~
{\cal P}_1=\sigma U
\bigg(
r^2 h'^2(1-\frac{2r_h}{r}(1-\frac{m}{r})+2h^2
\bigg),
\\
\label{vir2}
&&{\cal P}_2=\sigma
\bigg(
2h^2h'^2\big(-1+\frac{2m}{r}(2-\frac{r_h}{r} )\big)
-\frac{h^4}{r^2}(1-\frac{2r_h}{r} )
\bigg),
\end{eqnarray}
Setting $r_h=0$ in the above relations leads to a virial identity valid for
gravitating particle-like solutions.
Furthermore, a  virial relation for the nongravitating limit of
this model ($i.e.$ solutions of (\ref{eqs2}) in a fixed Minkowski spacetime background, no backreaction)
is found by taking $m=0,~r_h=0$ in (\ref{vir1}), (\ref{vir2}).

%%%%%%%%%%%%%%%%%%%%%%%%%%%%%%%%%%%%%%%%%%%%%%%%%%%%%%%%%%%%%%%
\subsection{The  global monopoles}
%%%%%%%%%%%%%%%%%%%%%%%%%%%%%%%%%%%%%%%%%%%%%%%%%%%%%%%%%%%%%%%
For $U(|\vec \phi|)=1$, $\lambda_2=0$,
the action (\ref{act}) corresponds to the usual global monopole model, with
%$V(|\vec \phi|)=\lambda (\eta^2-|\vec \phi|^2)^2$,
\begin{eqnarray}
\label{gm}
L_{m}= \partial _{\mu}\vec \phi \cdot\partial^{\mu}\vec \phi
 +\lambda(|\vec \phi|-\eta^2)^2,
\end{eqnarray}
%$ L_{m}= \partial _{\mu}\vec \phi \cdot\partial^{\mu}\vec \phi
 %+\lambda(|\vec \phi|-\eta^2)^2$,
 whose
 solutions were extensively studied in the literature.
The configurations with a regular origin are found with a usual shooting method by which one
adjusts the value of $h'(0)$,
integrates outward to large radius, and shoots for an asymptotic
boundary condition such that $h(r\to \infty) \to 1$. Such solutions represent global
monopoles of unit charge and are parametrized by $h'(0)$.
It turns out that above a critical value $\alpha =\alpha_{max}$ no such solutions can be found
\cite{Liebling:1999ke}, \cite{Liebling:1999bb}.
  As $\alpha$ is increased towards
the critical $\alpha_{max}$, the value of $h'(0)$ for which the static monopole solution is found decreases
toward zero. The critical solution represents the point at which the static monopole becomes
identical to de Sitter space in which $h(r) = 0$ and the symmetry-breaking potential
reduces to a cosmological constant \cite{Maison:1999pi}.
The picture for black hole sitting inside global monopoles is more complicated
\cite{Maison:1999ke}, \cite{Tamaki:2003kv}.
For values of $\alpha$ below some critical value, one finds black holes with
arbitrarily large radius.
Above this critical value, the branch of black holes bifurcates with the
Schwarzschild-de Sitter solution.

 However, a generic feature of all these solutions
 is that the kinetic term $\partial _{\mu}\vec \phi \cdot\partial^{\mu}\vec \phi $
 does not vanish at infinity, which leads to a divergent total mass\footnote{One
 can see from (\ref{eqs1}) that since $h\to 1$ for large enough $r$,
 then $m'\sim 2\alpha^2$ for solutions with $U=1$.}.
 The absence of finite mass solutions can also be seen from
the virial identity (\ref{vir1}),
since one can show that both ${\cal P}_0$ and ${\cal P}_1$ are strictly positive quantities.

 %%%%%%%%%%%%%%%%%%%%%%%%%%%%%%%%%%%%%%%%%%%%%%%%%%%%%%%%%%%%%%%
\subsection{The new model: {\it `regularised'} global monopoles}
%%%%%%%%%%%%%%%%%%%%%%%%%%%%%%%%%%%%%%%%%%%%%%%%%%%%%%%%%%%%%%%
 To cure this mass divergence, we consider in this work a nontrivial correction factor
 in front of the $\partial _{\mu}\vec \phi \cdot\partial^{\mu}\vec \phi $
 term,
 \begin{eqnarray}
\label{U}
U(|\vec \phi|)=(|\vec \phi|-\eta^2)^2,
\end{eqnarray}
where $|\vec \phi| \to \eta$ at infinity.
This expression of $U$ regularises the contribution of the kinetic term to the mass-energy.
However, the virial identity (\ref{vir1}) forbids again the existence
of finite mass solutions unless the Skyrme-like term is also included, $\lambda_2> 0$.
Technically, this is a consequence of the fact that the sign of ${\cal P}_2$ in (\ref{vir1})
is not fixed, and, in fact it becomes negative for large values of $r$.
Heuristically, similar to the Hopf or Skyrme models \cite{Radu:2008pp},
  the quartic term $(\partial _{\mu}\vec \phi \times \partial_{\nu}\vec \phi)^2$
provides the extra repealing force allowing  for finite mass to the solutions.

Also, by using the field equation
for $\vec \phi$,
one can show that no finite mass solutions are found in a truncated  model with
the Skyrme-like term only\footnote{To this end, one writes the equation for $h$ as $(\sigma N h' h^3)'=\sigma(2N h'^2h^2+h^4/r^2)$.
 Then by integrating the origin/event horizon to infinity, one can show that $h\equiv 0$.}.
Therefore,  we are forced to consider one more term in (\ref{L-gold})
in addition to $(\partial _{\mu}\vec \phi \times \partial^{\nu}\vec \phi)^2$.
This can be the kinetic term
$\partial _{\mu}\vec \phi \cdot\partial^{\mu}\vec \phi $
with an extra factor given by (\ref{U}) and/or a symmetry-breaking potential term.
%In what follows, we shall consider the physically more interesting case of
We have verified the existence of finite mass solutions of the equations (\ref{eqs1}) in a model with
 \begin{eqnarray}
\label{l4}
 L_{m}= (\partial _{\mu}\vec \phi \times \partial_{\nu}\vec \phi)^2
 +\lambda(|\vec \phi|-\eta^2)^2,
\end{eqnarray}
$i.e.$ with a quartic term only.
 However, we have found more interesting to keep
 all terms in (\ref{L-gold}), and to consider the general model with a correction factor
$U(|\vec \phi|)$ given by (\ref{U}).

One should remark that
although it may look unusual, the model (\ref{L-gold})
(with  $U(|\vec \phi|)$ given by
(\ref{U}), $V(|\vec \phi|)=\lambda(|\vec \phi|-\eta^2)^4$ and $\lambda_1=4\lambda_2=1$)
has some higher dimensional origin.
As discussed in  \cite{Tchrakian:1990ai},
the Lagrangian (\ref{L-gold}), considered on flat $\R^3$, arises from the
gauge decoupling limit ($i.e.$ no Yang-Mills fields) of the three dimensional $SO(3)$ gauged Higgs model
descended from the {\it second} ($p=2$) member of the Yang-Mills  hierarchy on $\R_3 \times S^{5}$.
%which was studied quantitatively in \cite{Kleihaus:1998kd}. In other words, the locally gauged version
%of the model employed here is the Yang-Mills--Higgs model of \cite{Kleihaus:1998kd}.
The resulting global symmetry breaking Goldstone model considered in a fixed four dimensional Minkowski
spacetime
background  admits finite energy solutions \cite{Paturyan:2005ik}
(see also Ref. \cite{Radu:2007zz} for
higher dimensional generalisations).
In this case,
it is straightforward to show that the corresponding energy density is bounded
from below by
\bea
\label{lb1}
\varrho=\frac{1}{4 \pi}
\vep_{ijk}\vep^{abc}\,
\left(\eta^2-|\vec\f|^2\right)\,\f_i^a\,\f_j^b\,\f_k^c
=\frac{1}{4 \pi}
 \vep_{ijk}\vep^{abc}\,\pa_i\,
\left[\left(\eta^2-\frac35|\vec\f|^2\right)
\,\f^a\,\f_j^b\,\f_k^c\right],
\label{totdiv}
\eea
(with $i,j=1,2,3$).
Then, as discussed in  \cite{Paturyan:2005ik}, the total
mass of the flat space solutions has a lower bound (which is never saturated),
$M\geq  Q$,  with $Q=\int d^3 x~\varrho=4/5$ a topological charge for our spherically symmetric
 configurations\footnote{The model (\ref{gm}) (with $V(|\vec \phi|)=0$) can also be
thought as resulting from the gauge decoupling limit of the {\it first}
member of the Yang-Mills  hierarchy on $\R_3 \times S^{1}$.
However, no lower bound  for their action (similar to that resulting from (\ref{lb1}))
is found in that case.}.
The locally gauged version of this model corresponds to a specific Yang-Mills--Higgs theory,
whose solutions were studied quantitatively in \cite{Kleihaus:1998kd}.
%which was studied quantitatively in \cite{Kleihaus:1998kd}. In other words, the locally gauged version
%of the model employed here is the Yang-Mills--Higgs model of \cite{Kleihaus:1998kd}.

However, for the purposes of this work, we are interested in this model mainly  because it provides
a simple toy model admiting finite mass, black hole solutions with a
symmetry breaking scalar field outside the horizon, carying also a topological charge.

%%%%%%%%%%%%%%%%%%%%%%%%%%%%%%%%%%%%%%%%%%%%%%%%%%%%%%%%
 \subsection{Boundary values and asymptotic behaviour}
 %%%%%%%%%%%%%%%%%%%%%%%%%%%%%%%%%%%%%%%%%%%%%%%%%%%%%%%
We start by noticing that by using a suitable redefinition of $\lambda_1, \lambda_2$
together with a rescaling of the radial coordinate,
 one can always take $\lambda_1=4\lambda_2=1$ without any loss of generality.
 Also,
 to simplify
 the general picture, we set $V(|\vec \phi|)=0$ in what follows
 (although we could confirm that finite mass solutions
 exist also for a nonvanishing scalar potential).
 This leaves us with a single essential parameter of the problem, $\alpha$, which, for a given $G$,
 is fixed by the $v.e.v.$ of the Goldstone field.

The asymptotic form of the  functions $m,\sigma,h$ can be systematically constructed in both regions,
near the  event horizon/origin  and for $r \to \infty$.
The nonextremal solutions possess the following expansion near the event horizon:
\begin{eqnarray}
\label{eh}
 &&
 m(r)=\frac{r_h}{2}+m_1(r-r_h)+O(r-r_h)^2,~~\sigma(r)=\sigma_h+\sigma_1(r-r_h)+O(r-r_h)^2,
 \\
 \nonumber
&&
h(r)=h_0+h_1(r-r_h)+O(r-r_h)^2.
\end{eqnarray}
For a given event horizon radius $r_h$, the essential parameters characterizing the event horizon
are $h_0$ and $\sigma_h>0$, which fix all higher order coefficients in (\ref{eh}).
(These constants are related in a complicated way to the parameters $M,c_1$ of the
solutions in the far field expansion (\ref{expansion-infty}).)
One finds $e.g.$ for the lowest order terms
\begin{eqnarray}
 \nonumber
 &&m_1=
 \frac{\alpha^2 h_0^2}{r_h^2}
 \left(
 h_0^2+2(1-h_0^2)^2r_h^2
 \right),~~
 h_1=\frac{4h_0}{2(1-2m_1)r_h}
 \frac{\big( (1-h_0^2)^2r_h^2+h_0^2(1-2(1-h_0^2)r_h^2)\big)}
 {
 \big(2h_0^2+ (1-h_0^2)^2r_h^2 \big)
 },
 \\
 \label{cond1}
&&
 \sigma_1=
 2\alpha^2 \sigma_h\frac{h_1^2}{r_h}
\left(
 2h_0^2+ (1-h_0^2)^2r_h^2
 \right).
\end{eqnarray}
The Hawking temperature and the entropy of the black holes are given by
\begin{eqnarray}
\label{THS}
T_H= \frac{1}{4 \pi} \sigma(r_h) N'(r_h) ,~~S=\frac{A_H}{4G},~~{\rm with}~~N'(r_h)=\frac{1}{r_h}(1-2m'(r_h))
~~{\rm and}~~A_H= 4 \pi r_h^2.
\end{eqnarray}

As we shall see, in the limit of   zero event horizon radius of the black holes,
the solutions describe
particle-like globally regular solitons.
$r=0$ is in this case a regular origin, the
 corresponding  approximate solution
close to that point
 being
\begin{eqnarray}
\label{expansion-0}
&&
h(r)= \bar h_1r+h_3 r^3+O(r^5),~~
m(r)= \alpha^2 \bar h_1^2(1+\bar h_1^2) r^3+O(r^5),~~
\\
\nonumber
&&
\sigma(r)=\sigma_0+\alpha^2 \sigma_0\bar h_1^2(1+2 \bar h_1^2) r^2+O(r^4),
\end{eqnarray}
where
\begin{eqnarray}
\label{h3}
h_3=\frac{\bar h_1^3\left(-1+\alpha^2(3+6\bar h_1^2+2\bar h_1^4)\right)}{5(1+2\bar h_1^2)},
\end{eqnarray}
with two free parameters $\bar h_1=h'(0)$ and $\sigma_0$.
%As $r\to \infty$, these configurations approach the asymptotic
%solution (\ref{expansion-infty}).

We assume that the spacetime is asymptotically flat, which leads to the
following expansion as $r\to \infty$
\begin{eqnarray}
\label{expansion-infty}
h(r)=1+c e^{-2r}-\frac{1}{4r^2}+\dots,~m(r)=M-\frac{\alpha^2}{4r}+\dots,~~
\sigma(r)=1-\frac{\alpha^2}{6r^6}+\dots~,
\end{eqnarray}
in terms of two free parameters $M$, $c_1$, with $M$ the mass of the solutions.

%%%%%%%%%%%%%%%%%%%%%%%%%%%%%%%%%%%%%%%%%%%%%%%%%%%%%%%%%%%%%%%
\section{Numerical results}
%%%%%%%%%%%%%%%%%%%%%%%%%%%%%%%%%%%%%%%%%%%%%%%%%%%%%%%%%%%%%%%

%%%%%%%%%%%%%%%%%%%%%%%%%%%%%%%%%%%%%%%%%%%%%%%%%%%%%%%%%%%%%%%
\subsection{The solitons}
%%%%%%%%%%%%%%%%%%%%%%%%%%%%%%%%%%%%%%%%%%%%%%%%%%%%%%%%%%%%%%%
Since the black holes of our model smoothly emerge from
the particle-like solutions, we start by discussing this
 limit.

 The solitons are the gravitating generalisation of the flat space
solutions considered in \cite{Paturyan:2005ik}.  This limit  is recovered
for $\alpha=0$, $N(r)=\sigma(r)=1$.
The backreaction is included by slowly increasing the value of $\alpha$.
For a given $\alpha$, the solutions may exist for discrete values of
$h'(0)$
 (which is the  shooting parameter of the problem).

We follow the usual approach and, by using a
standard ordinary differential equation solver\footnote{Some of the solutions were also
constructed by using an isotropic
coordinate system, with $f_2=f_1 r^2$ in the general ansatz (\ref{metr-sph}).
In that case we have employed a different  solver \cite{schoen} which involves a Newton-Raphson method for boundary value
ordinary differential equations. This approach also allows us to extend our consideration
to a more general case of the solutions with axial symmetry and higher values of the winding number 
(this study will be reported elsewhere).}, we evaluate the initial conditions  (\ref{expansion-0})
at $r = 10^{-6}$, adjusting for the fixed shooting parameter
and integrating towards $r\to \infty$.
The integration
stops when the  asymptotic limit (\ref{expansion-infty})  is reached  with
reasonable accuracy.

Although solutions with nodes in $h(r)$ do also exist, we discuss in what follows the
configurations with a monotonic behaviour of the scalar function only.
As one can see from the field equations (\ref{eqs1}), the metric functions
$m,\sigma$ are also strictly increasing with $r$.
The Figure 1 shows the profiles of three solutions with
different values of $\alpha$.

As expected, the solutions exist for a finite range of $\alpha$,
with $\alpha<\alpha_{max}\simeq 1.545$.
%As expected, increasing $\alpha$ leads to
The picture can be summarize as follows:
when  $\alpha$
increases, the dimensionless mass 
%
 %%%%%%%%%%%%%%%%%%%%%%%%%%%%%%%%%%%%%%%%%%%%%%%%%%%%%%%%%%%%%%%%%%%%%%%%%%%
\setlength{\unitlength}{1cm}
\begin{picture}(8,6)
\put(-0.7,0.0){\epsfig{file=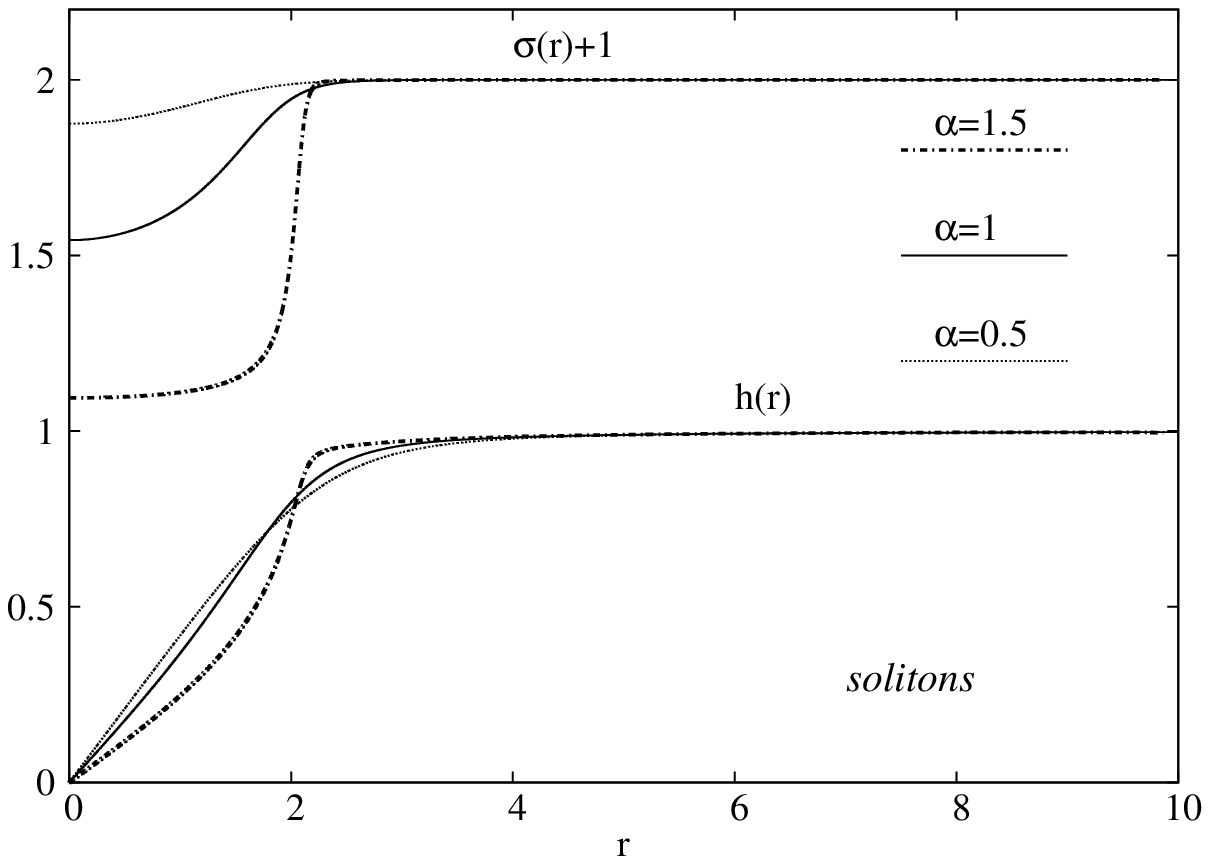,width=8cm}}
\put(7.8,0.0){\epsfig{file=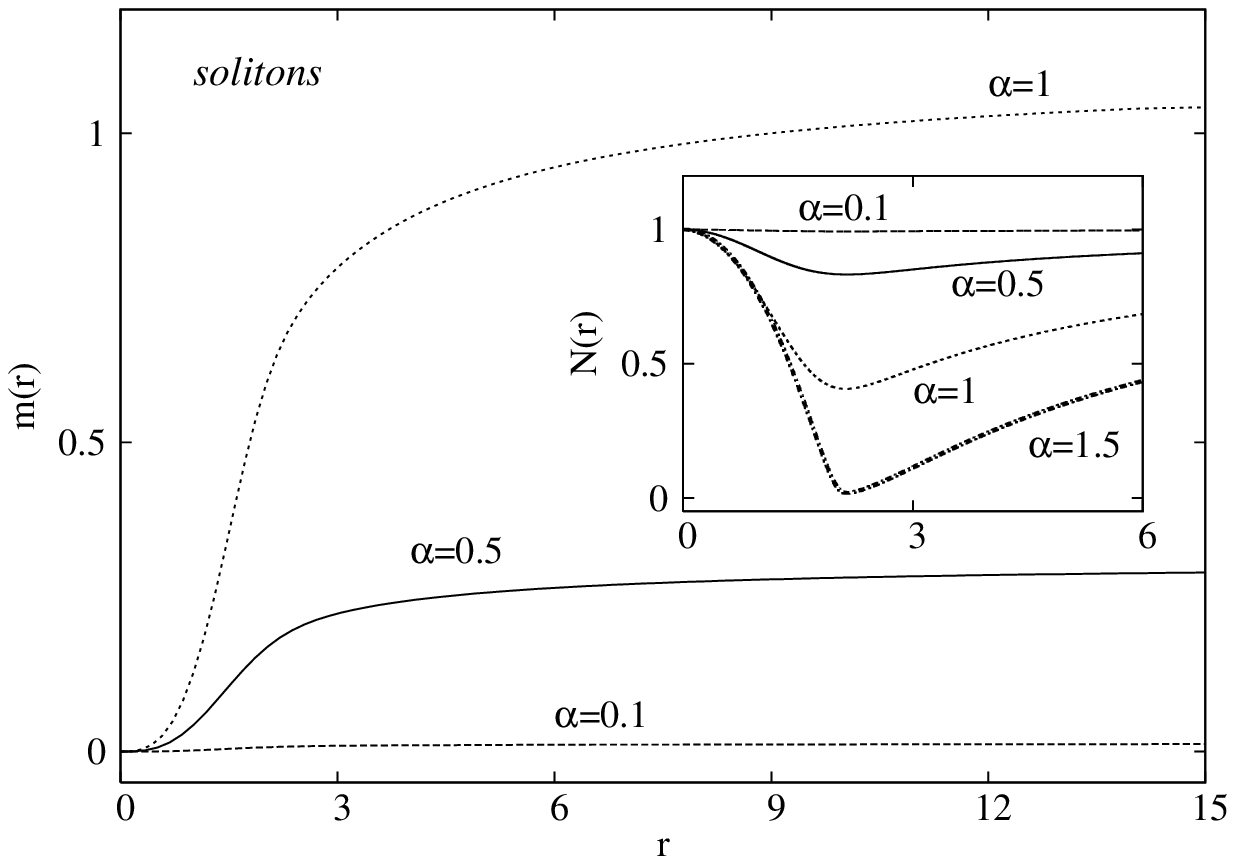,width=8cm}}
\end{picture}
\\
\\
{\small {\bf Figure 1.}
The profiles of  scalar soliton solutions
are shown for several different values of the coupling constant $\alpha$. }
\vspace{0.5cm}
%%%%%%%%%%%%%%%%%%%%%%%%%%%%%%%%%%%%%%%%%%%%%%%%%%%%%%%%%%%%%%%%%%%%%%%%%%%
\\
%%%%%%%%%%%%%%%%%%%%%%%%%%%%%%%%%%%%%%%%%%%%%%%%%%%%%%%%%%%%%%%%%%%%%%%%%%%
\setlength{\unitlength}{1cm}
\begin{picture}(8,6)
\put(-0.7,0.0){\epsfig{file=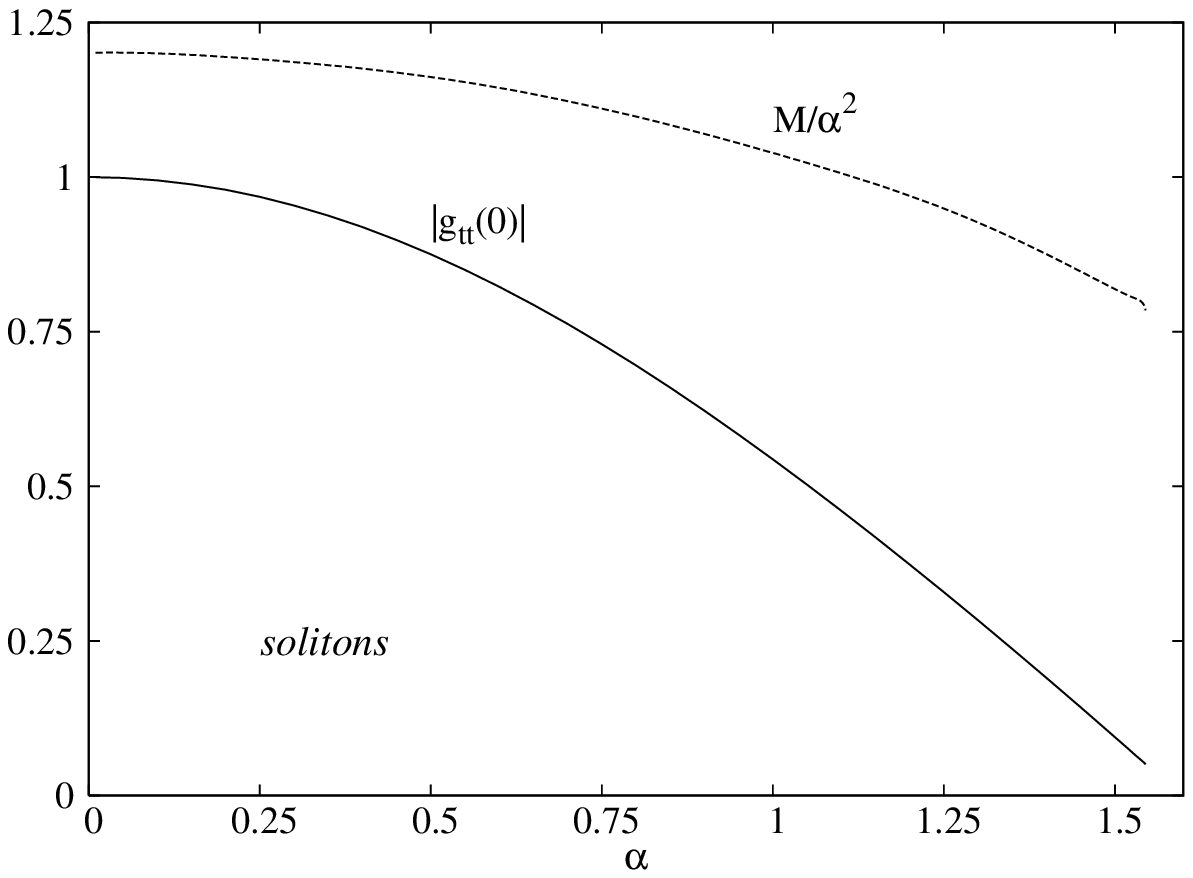,width=8cm}}
\put(7.9,0.0){\epsfig{file=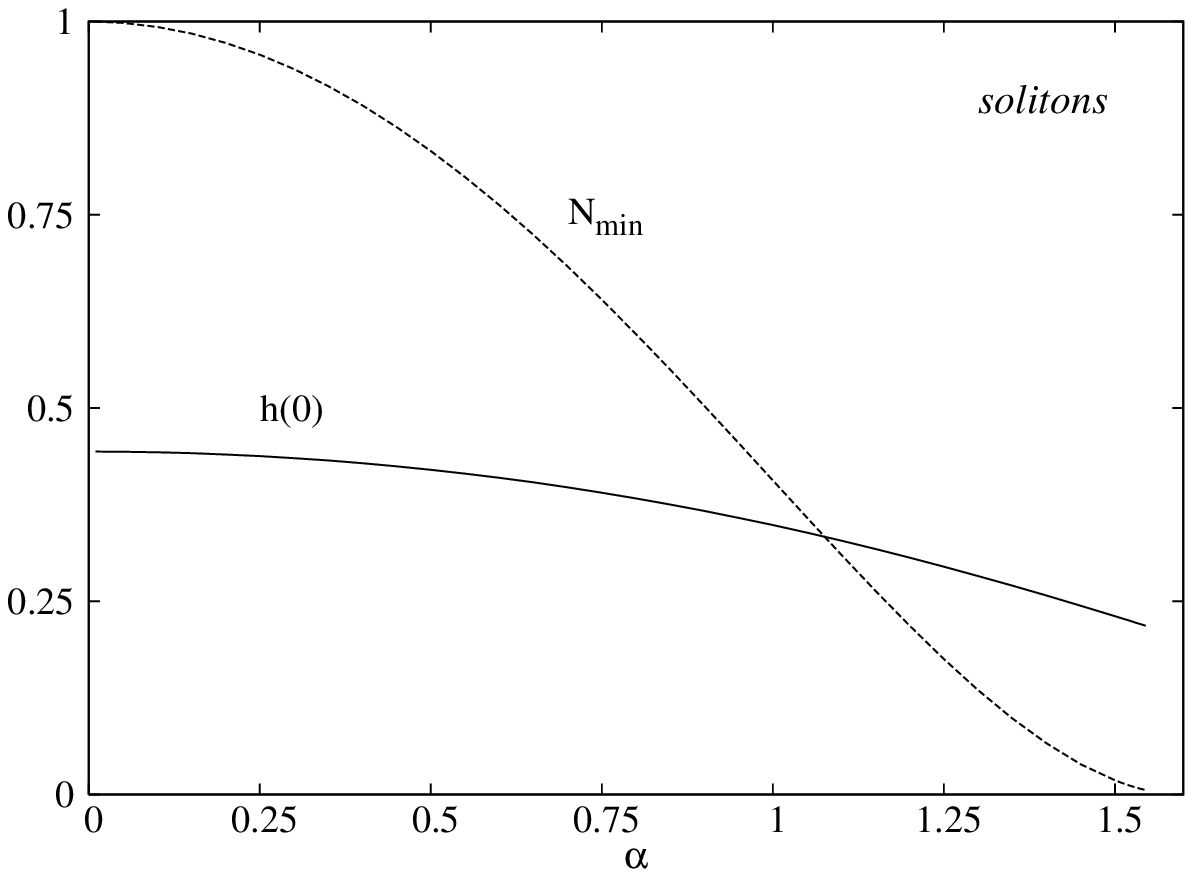,width=8cm}}
\end{picture}
\\
\\
{\small {\bf Figure 2.}
Some features of the scalar soliton solutions are shown as a function of the coupling parameter $\alpha$. }
\vspace{0.5cm}
%%%%%%%%%%%%%%%%%%%%%%%%%%%%%%%%%%%%%%%%%%%%%%%%%%%%%%%%%%%%%%%%%%%%%%%%%%%
\\
parameter $M/\alpha^2$ decreases, as well as the value $\sigma(0)$.
The minimum $N_m$
of the function $N(r)$ also decreases, as indicated in Figures 1,2.
This minimum  becomes more pronounced for larger $\alpha$, and finally,
a horizon is found for  $\alpha=\alpha_{max}$ and some finite value of  $r=r_c\simeq 2.15$.

We have noticed that, within the numerical errors, $N'(r_c)=0$,
$i.e.$ the function $N(r)$ has a double zero at $r=r_c$.
Also, the proper distance $\ell=\int_{0}^r dr/\sqrt{N(r)}$ diverges for $r\to r_c$.
As a result, the spatial geometry on the hypersurface
$t =const.$ develops an infinite throat separating the interior region with a
smooth origin and non-trivial scalar field, from the exterior region which corresponds
to a finite mass, extremal black hole with scalar hair.
\\
%%%%%%%%%%%%%%%%%%%%%%%%%%%%%%%%%%%%%%%%%%%%%%%%%%%%%%%%%%%%%%%%%%%%%%%%%%%
\setlength{\unitlength}{1cm}
\begin{picture}(8,6)
\put(-0.1,0.0){\epsfig{file=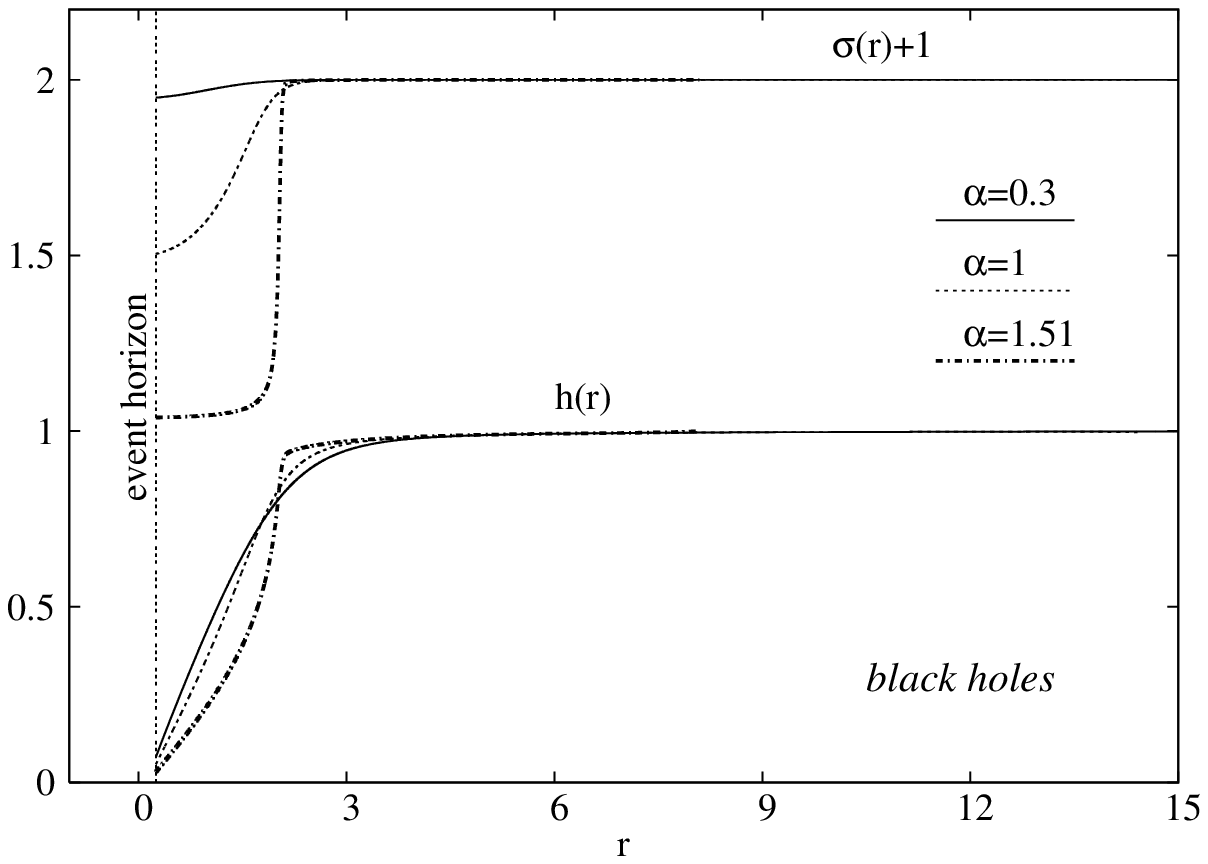,width=8cm}}
\put(8,0.0){\epsfig{file=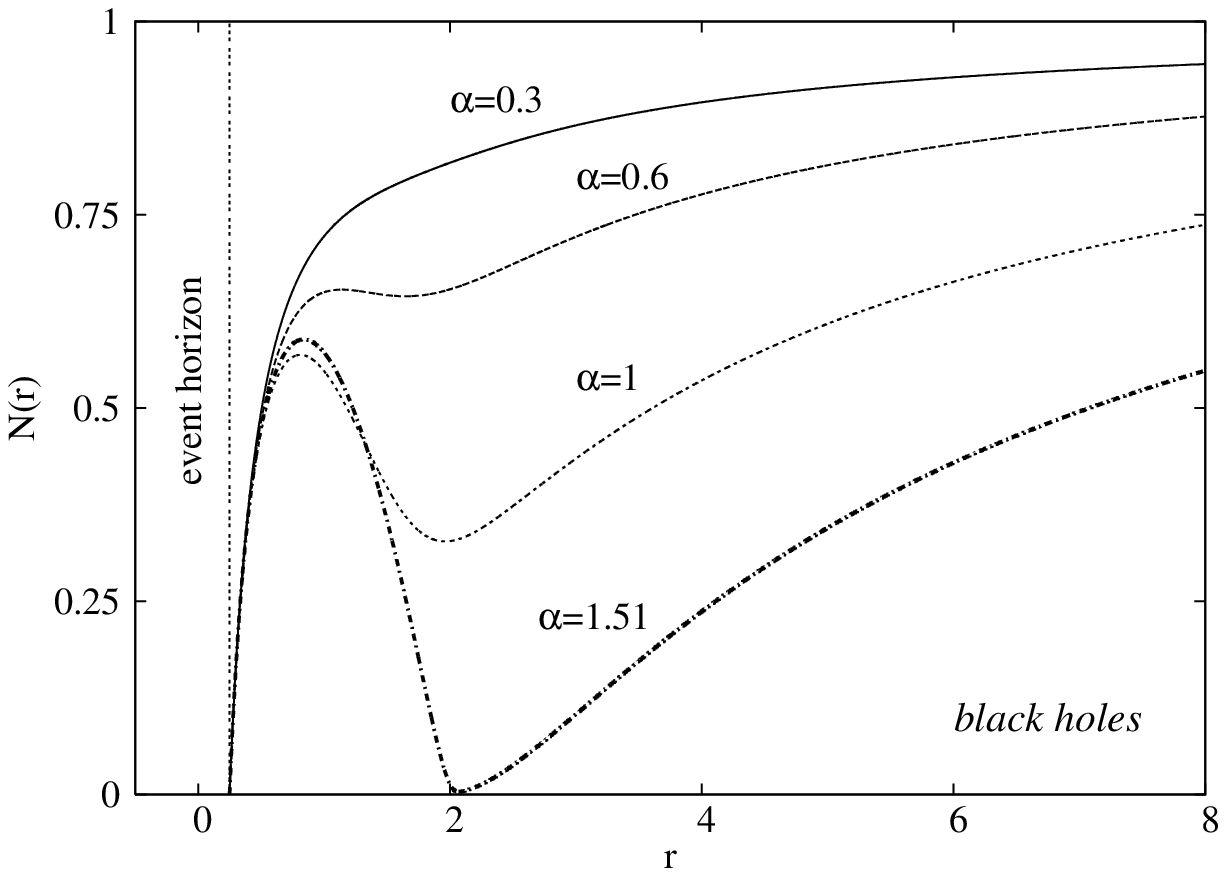,width=8cm}}
\end{picture}
\\
\\
{\small {\bf Figure 3.}
The profiles of typical black hole solutions with different values of $\alpha$
and the same event horizon radius are shown  as a function of the radial coordinate. }
\vspace{0.5cm}
%%%%%%%%%%%%%%%%%%%%%%%%%%%%%%%%%%%%%%%%%%%%%%%%%%%%%%%%%%%%%%%%%%%%%%%%%%%

It may be interesting to remark that this strongly contrasts with the picture found in \cite{Maison:1999pi} for
the global monopoles.
There, a de Sitter spacetime is approached for a maximal value of $\alpha$, with $h'(0)\to 0$ in that limit.
However,
for the model under consideration in this work,  the picture  is qualitatively
  similar to that valid for gravitating 't Hooft-Polyakov
monopoles \cite{gmono}.
In that case, the metric in the exterior region is that of an extremal Reissner-Nordst\"om black hole.
The near horizon geometry of the exterior configuration here is described by
the exact solution (\ref{att1}) with $h_0\simeq 0.92$.

%%%%%%%%%%%%%%%%%%%%%%%%%%%%%%%%%%%%%%%%%%%%%%%%%%%%%%%%%%%%%%%
\subsection{The black holes}
%%%%%%%%%%%%%%%%%%%%%%%%%%%%%%%%%%%%%%%%%%%%%%%%%%%%%%%%%%%%%%%

According to the standard arguments, one can expect black hole generalisations of the regular
configurations to exist at least for small values of the horizon radius $r_h$.
This is indeed confirmed
by
the numerical analysis.
In our approach, we have restricted our integration
to the (physically more relevant) region outside
of the horizon, $r \geq r_h$.
Given $(r_h,\alpha)$, the black hole solutions may exist for a set of discrete values
 of the
scalar field on the horizon,
$h_0$.
Similar to the soliton case, we restrict our study to solutions with no nodes in $h(r)$.
The profiles of several solutions with the same event horizon radius
are shown in Figure 3, for several values of $\alpha$.

Starting from
a regular solution with a given $\alpha$ and increasing the event horizon radius,
we find a  branch of black hole solutions.
For $r_h \ll 1$, the solutions resemble small black holes
sitting in the center of the regular lumps, the latter being almost unaffected
for $r \gg r_h$ by the presence of the black hole.
The Hawking temperature decreases along this branch while the mass and
the value of the scalar field on the horizon increase.

The issue of the limiting behaviour of this branch of solutions
is more complicated and depends crucially on the value of
$\alpha$.
For $\alpha<\alpha_c=3\sqrt{3}/4$,
we could construct solutions with very large value of $r_h$ and they are likely to exist for
arbitrary values of the event horizon radius.
The value at the horizon of the scalar field increases with $r_h$,
approaching 
\\
%%%%%%%%%%%%%%%%%%%%%%%%%%%%%%%%%%%%%%%%%%%%%%%%%%%%%%%%%%%%%%%%%%%%%%%%%%%
\setlength{\unitlength}{1cm}
\begin{picture}(8,6)
\put(0.,0.0){\epsfig{file=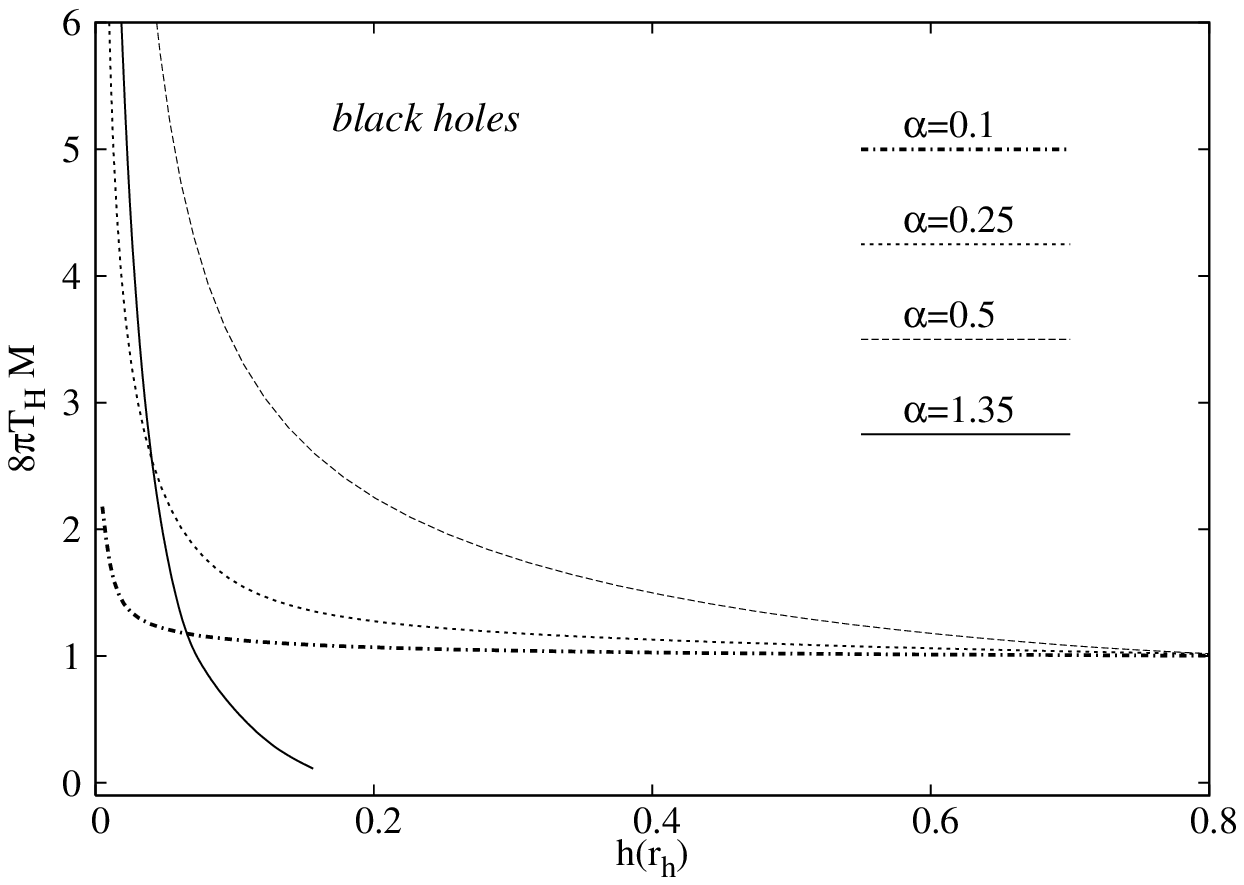,width=8cm}}
\put(8,0.0){\epsfig{file=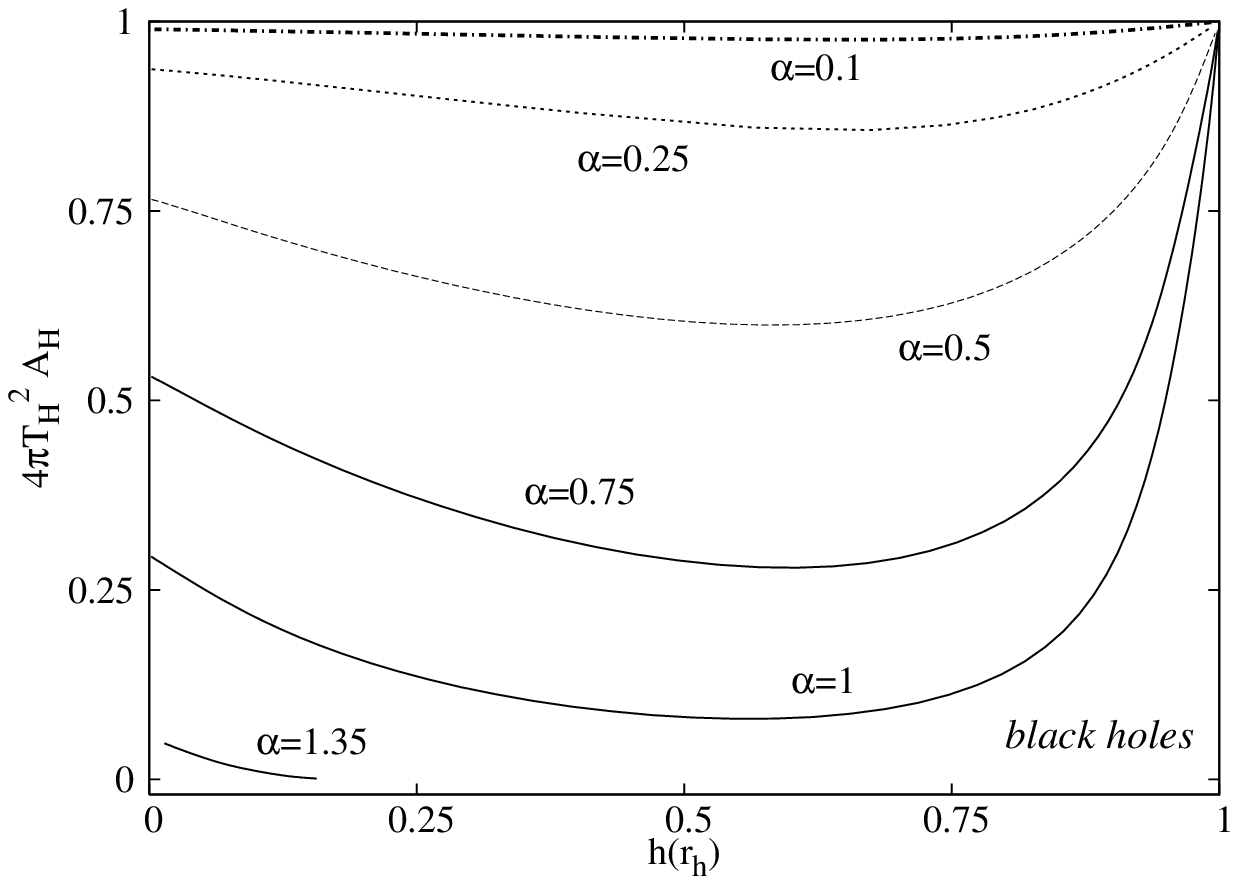,width=8cm}}
\end{picture}
\\
\\
{\small {\bf Figure 4.}
The (suitably normalized-) dimensionless quantities $T_H M$ and $T_H^2 A_H$
are shown as functions
of the scalar field on the horizon $h(r_h)$,
 for several fixed values of the parameter $\alpha$.
 The control parameter here is the event horizon radius $r_h$,
 with
 $h(r_h)=0$ and $h(r_h)=1$ corresponding
to the soliton and the Schwarzschild black hole limits, respectively.}
\vspace{0.5cm}
%%%%%%%%%%%%%%%%%%%%%%%%%%%%%%%%%%%%%%%%%%%%%%%%%%%%%%%%%%%%%%%%%%%%%%%%%%%
\\
%%%%%%%%%%%%%%%%%%%%%%%%%%%%%%%%%%%%%%%%%%%%%%%%%%%%%%%%%%%%%%%%%%%%%%%%%%%
\setlength{\unitlength}{1cm}
\begin{picture}(8,6)
\put(-0.2,0.0){\epsfig{file=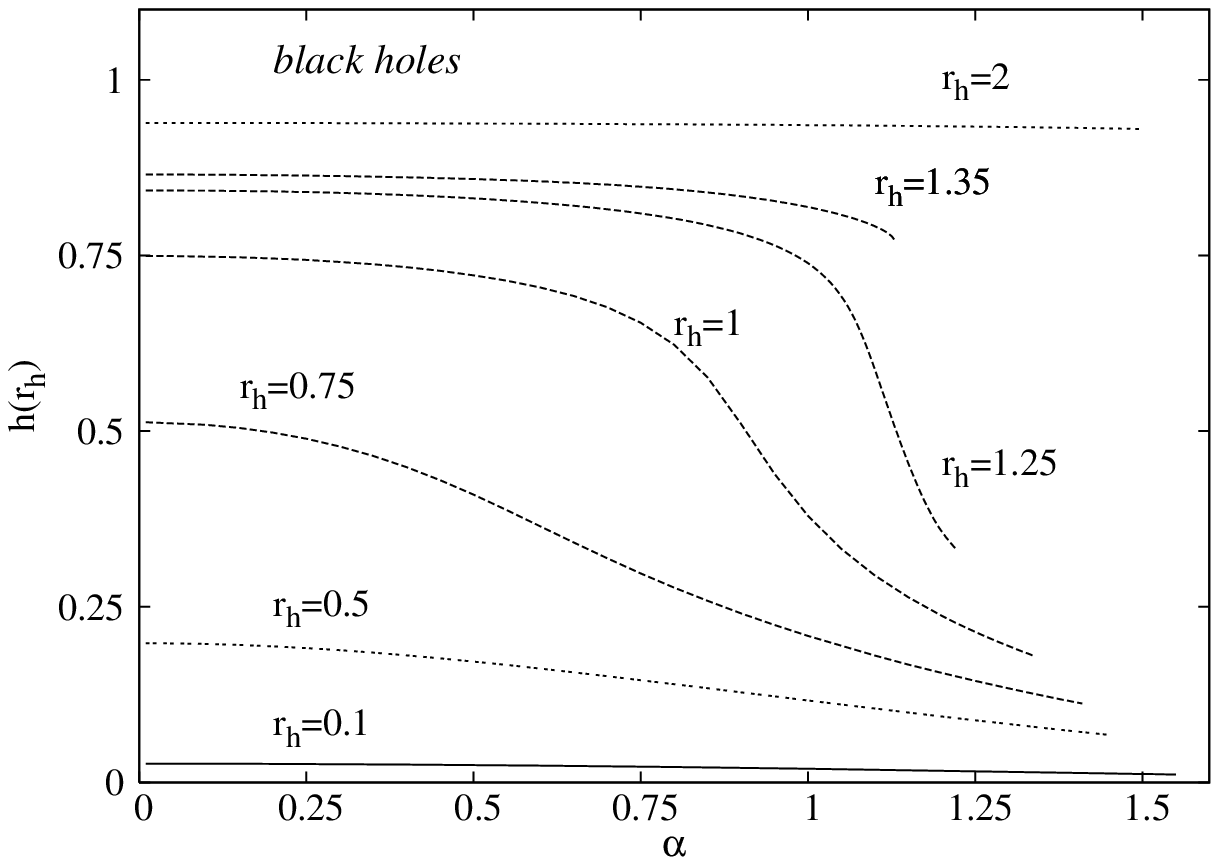,width=8cm}}
\put(8.2,0.0){\epsfig{file=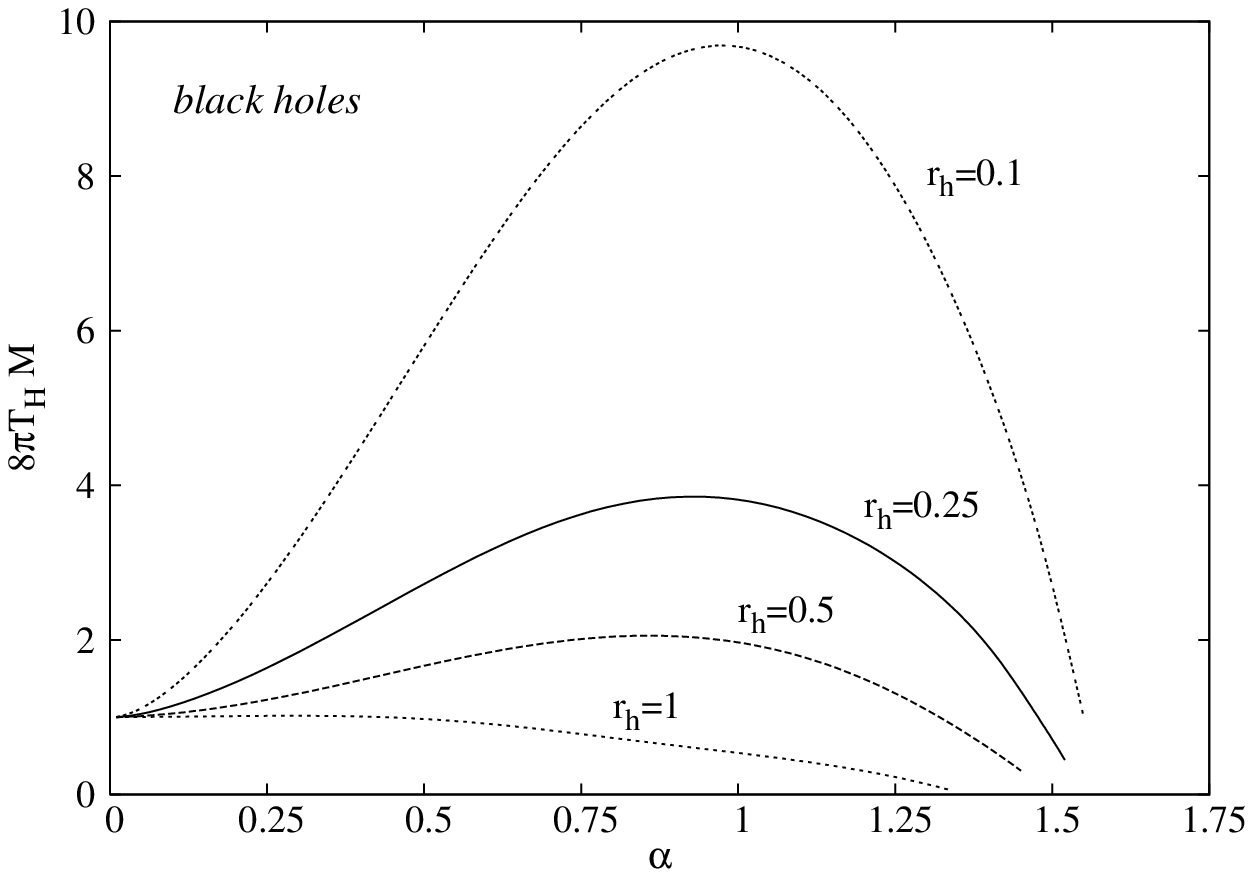,width=8cm}}
\end{picture}
\\
\\
{\small {\bf Figure 5.}
Some features of the black hole solutions are shown as functions
of $\alpha$,
 for several fixed values of the event horizon radius. }
\vspace{0.5cm}
%%%%%%%%%%%%%%%%%%%%%%%%%%%%%%%%%%%%%%%%%%%%%%%%%%%%%%%%%%%%%%%%%%%%%%%%%%%
\\
asymptotically the unit value.
As seen in Figure 4, the large black holes are essentially Schwarzschild solutions,
with a small scalar field outside the horizon.
The intrinsic parameter there is the value of the scalar field on the horizon, with $h(r_h)=0$ reached as $r_h \to 0$
and $h(r_h)$ very close to one for large values of the the event horizon radius; also, we have
 normalized
$T_H M$ and $T_H^2 A_H$ such that the unit value is approached for a Schwarzschild black hole.

The picture is different for $\alpha_c<\alpha<\alpha_{max}$ (for example for $\alpha=1.35$ in Figure 4),
in which case we notice the following pattern.
As $r_h$ increases, the metric function $N(r)$ starts to develop
a second minimum, and, for some critical value of $r_h$,
one finds $N(r_c)=N'(r_c)=0$,  with $r_c>r_h$ depending on $\alpha$.
Therefore an extremal horizon occurs in the outside region, which is the limiting configuration
of this set of solutions (thus the range of $r_h$ is bounded for $\alpha_c<\alpha<\alpha_{max}$).
This horizon is regular, as found by computing some curvature invariants,
with both $\sigma(r_c)$ and $h(r_c)$ taking values different from one there.
Thus these black holes develop an infinite throat and become gravitationally closed.

It is instructive to consider also a fixed value of $r_h$
and to vary the value of $\alpha$.
As one can see in Figure 5, the solutions with $\alpha=0$ correspond in this case to
"regularised" global monopoles in a fixed Schwarzschild background
($e.g.$ one finds $8 \pi T_H M \to 1$ as $\alpha\to 0$).
As $\alpha$ increases, the geometry of the spacetime deviates from the
 Schwarzschild one. Then, for a critical value of $\alpha$,
 the metric function $N(r)$, apart from a simple zero at $r = r_h$,
develops again a double zero at some $r_c>r_h$.
Again, a similar picture is found for black holes with large enough values of $\alpha$
in the gravitating Georgi-Glashow
model, see $e.g.$ the review work \cite{Volkov:1998cc}.

%%%%%%%%%%%%%%%%%%%%%%%%%%%%%%%%%%%%%%%%%%%%%%%%%%%%%%%%%%%%%%%
\subsection{Extremal black holes.
An AdS$_2\times S^2$ exact solution}
%%%%%%%%%%%%%%%%%%%%%%%%%%%%%%%%%%%%%%%%%%%%%%%%%%%%%%%%%%%%%%%

 The near horizon expansion of the extremal black holes is more constrained,
since the metric function $N(r)$ has a double zero at the horizon,
$N(r)=N_2(r-r_h)^2+O(r-r_h)^3,$ while the expansion for $\sigma(r)$
and $h(r)$ is still given by (\ref{eh}), with
\begin{eqnarray}
\label{rel2}
&&N_2=\frac{(1-h_0^2)(1-3h_0^2)^2}{h_0^2(1+h_0^2)},~~
\sigma_1=\frac{4(1+h_0^2)(5h_0^2-1)
\left(
(h_0^2-1)(1-3h_0^2)
\right)^{5/2}}
{h_0(3-7h_0^2+17 h_0^4-21 h_0^6)^2}\sigma_h,~~
\\
\nonumber
&&h_1=\frac{2(1+h_0^2)\left(
(h_0^2-1)(1-3h_0^2)
\right)^{3/2}}{-3+7h_0^2-17h_0^4+21 h_0^6}.
\end{eqnarray}
Moreover, the event horizon radius of an extremal configuration is also fixed,
$r_h= \frac{h_0}{\sqrt{(h_0^2-1)(1-3 h_0^2)}}$.
The parameter $h_0=h(r_h)$ is determined by the coupling constant $\alpha$
as a solution of the equation $\sqrt{2}h_0\sqrt{1-h_0^4}=1/\alpha$.

The occurrence of extremal
black holes in our model is also suggested by the existence of the following  exact solution
of the Einstein-Goldstone equations (\ref{Eeq}), (\ref{Geq}),
corresponding to an $AdS_2 \times S^2$ spacetime with a constant scalar field magnitude:
 \begin{eqnarray}
\label{att1}
ds^2=v_1(\frac{dr^2}{r^2}-r^2 dt^2)+v_2(d\theta^2 + \sin^2 \theta d\varphi^2),~~h(r)=h_0,
\end{eqnarray}
where
\begin{eqnarray}
\label{att2}
 v_1=\frac{h_0^2(1+h_0^2)}{(3 h_0^2-1)^2(1-h_0^2)},~~
  v_2=\frac{h_0^2 }{(3 h_0^2-1) (1-h_0^2)},~~
  \alpha=  \frac{1}{\sqrt{2}h_0 \sqrt{1-h_0^4}} ~.
\end{eqnarray}
As seen from (\ref{rel2}), this solution describes the
neighborhood of the event horizon of an extremal black hole.
(The far field expression
of an extremal solution is still given by (\ref{expansion-infty}).)
The configuration (\ref{att1}) provides also an analytical explanation for some features
revealed in the numerical analysis.
For example, one can see that the range of $h_0$ is restricted, $1/\sqrt{3}<h_0<1$, while
 $3\sqrt{3}/{4}<\alpha <\infty$.
 (Also, as $h_0$ approaches the limiting values, the sizes of the $AdS_2$ and $S^2$
 parts of the metric diverge.)
 Then it is clear that extremal black holes may exist for
 $\alpha >3\sqrt{3}/{4}$ only.
It would be interesting to consider these solutions in the context of the attractor mechanism
and to compute their entropy function.

 One can verify that no  $AdS_2 \times S^2$
 solutions are found in the global monopole model (\ref{gm}),
 which is consistent with the absence of extremal black holes for the numerical solutions
 discussed in \cite{Liebling:1999ke}, \cite{Maison:1999ke}.

%%%%%%%%%%%%%%%%%%%%%%%%%%%%%%%%%%%%%%
\section{Further remarks}
%%%%%%%%%%%%%%%%%%%%%%%%%%%%%%%%%%%%%

The main purpose of this work was to present finite mass black hole
and soliton solutions in a theory with scalar fields featuring a
spontaneously broken symmetry. To this end we have used a specific Goldstone
model described by an $O(3)$ isovector scalar field, originally proposed in \cite{Tchrakian:1990ai}.

Not entirely surprisingly, it turns out that a number of basic feature of our solutions
are rather similar to those of the well-known gravitating gauged monopoles,
and not to the usual global monopoles solutions of the model (\ref{gm}).
What the present (global-Goldstone) monopole has in common with the local gauged ('t~Hooft-Polyakov) monopole
is the topological charge and lower bound, both of which are absent in the usual global monopole. In both cases the
topology is encoded in the scalar iso-multiplet. This may indicate that, also in the gravitating case,
some basic properties of the gauged monopole can be attributed to the scalar fields.

That our gravitating Goldtsone solutions are akin to gravitating 't~Hooft-Polyakov monopoles~\cite{gmono},
and distinct from gravitating Skyrmions \cite{gskyrm} is expected. This is because like the former~\cite{gmono}, our
solutions decay as $monopoles$ unlike the Skyrmions which decay like instantons, $i.e.$ at a faster rate as {\it pure gauge}.
A very simple manifestation of this feature is the different behaviours of the radial functions $h(r)$ and $w(r)$ describing
$(a)$ the Higgs field of a spherically symmetric monopole, and respectively $(b)$ the gauge field of a spherically symmetric
instanton. The boundary values of these functions between $[r=0\,,\,r=\infty]$ are $[h(0)=0\,,\,h(\infty)=1]$ and
$[w(0)=\pm 1\,,\,w(\infty)=\mp 1]$ respectively. It is also known~\cite{nonlin} that the dynamics of the unit charge
soliton of a $O(D+1)$ (Skyrme) sigma model on $\R^D$, described by the $chiral$ function $f(r)$
is identical to that of the corresponding Yang-Mills (YM) instanton form factor $w(r)$, $via$, $w(r)=\cos f$.
Thus, the boundary values
of the chiral function $f(r)$ of a Skyrmion, $[f(0)=\pi\, {\rm or}\, 0\,,\,f(\infty)=0\, {\rm or}\, \pi]$, are instantonic.
It is therefore not surprising that with gravitating Goldstone solutions we encounter extremal black holes, which in the case
of gravitating Skyrmions these are absent, just as the case is for the Einstein-YM (sphaleron) black holes in \cite{89}.

Having said this however, the Skyrmion is a topologically stable soliton like the Goldstone soliton, and unlike the YM sphaleron.
In this respect, the gravitating Skyrmion is more akin to our Goldstone solutions and different from the Bartnik-McKinnon
solution \cite{Bartnik:1988am} of the  Einstein-YM model,
as shown to be stable in \cite{Heusler:1992av}.
Also, although the dominant energy conditions is satisfied by the model in this work, similar
to the Skyrme theory, it involves nonrenormalizable interactions.

Also, similar to the case of Einstein-Skyrme theory,
we could  show that the specific gravitating Goldtsone model in this work has also black hole solutions stable against
linear fluctuations.
% (this was likely since we have found nodeless solutions).
In examining time-dependent fluctuations around the solutions in Section 3,
all field variables are written as the sum of the static equilibrium solution whose stability
we are investigating and a time dependent perturbation. By following the standard methods,
we derive linearized equations for $\delta \sigma(r, t)$, $\delta N(r, t)$
and $\delta h(r, t)$.   The linearized equations imply that
$\delta \sigma(r, t)$, $\delta N(r, t)$
are determined by $\delta h(r, t)$.
For an harmonic time dependence
$e^{-i\Omega t}$, the linearized system of the matter sector implies a standard Schr\"odinger equation
 \begin{eqnarray}
\label{Seq}
\left \{-\frac{d^2}{d\rho^2} +U(\rho) \right \}\beta(\rho)=\Omega^2 \beta(\rho) ,
\end{eqnarray}
where  $\beta = \delta h/g $ (with $g$ a strictly positive
function of the unperturbed variables) and a new  radial coordinate is introduced, $d/d\rho = N\sigma d/dr$.
 The expression of the potential is very complicated and we shall not give it here.
However, $U(\rho)$ is regular everywhere, with $U(r_h)=0$ and $U(\infty)=4$. It follows that Eq. (\ref{Seq})
 will have no bound
states if the potential $U(\rho)$ is everywhere greater than the lower of its two asymptotic values $i.e.$ $U(\rho)> 0$.
Indeed, this condition was satisfied by some of our black hole solutions.

Perhaps the most unusual feature of the model in this work
is the existence of extremal black hole configurations, even in the absence of gauge fields.
Moreover, by using the approach in \cite{Zaslavsky:1992sp}, \cite{Doneva:2011gx} one can show
%after a straightforward but cumbersome computation
the first law of thermodynamics for our black hole solutions reads
%\begin{eqnarray}
$dM=T_H dS  .$
%\end{eqnarray}
Thus there is no work term associated with the scalar field, which partially can be attributed
to the $1/r^2$ decay of the scalar field $\phi$ at infinity, see (\ref{expansion-infty}).
(Note that the scalar charge $\Sigma$, as defined from the large $r$ expression $\phi \sim \phi_{\infty}+\Sigma/r$,
may enter the first law  \cite{Gibbons:1996af}). The same form of the first law
is found for black holes with Skyrme hair \cite{Zaslavsky:1992sp}, \cite{Heusler:1993cj}; however, no extremal configurations
exist in that case.
The existence of extremal black holes for the model in this work,
viewed as a balance of two different charges, may look puzzling,
since the only charge we can define on the scalar sector is a topological one.
However, this can be understood in analogy with the magnetically charged Reissner-Nordstr\"om black hole
embedded in a non-Abelian theory.
In that case, the magnetic charge is quantised and, accordingly, there is no work term
associated with it in the first law, despite the existence of extremal black holes.
Similarly, the solutions considered in this work have unit 'magnetic' charge,
which has a topological origin and likewise cannot enter the first law.

As a direction for future work, it would be interesting to study the case of solutions with a
nonvanishing scalar symmetry-breaking potential. In the presence of gauge fields, the inclusion of this term
is known to change some properties of gravitating monopoles~\cite{gmono} drastically. Here, the same would be
expected of the gravitating Goldstone, since that is also a monopole theory.

Also, working in a flat spacetime background,  Ref. \cite{Paturyan:2005ik}
has given numerical evidence for the existence of axially symmetric generalisations of the
 spherically symmetric  solutions of the model in this work.
  These are
multisolitons with topological charge $n>1$, and unstable soliton-antisoliton pairs with zero
topological charge. By employing methods similar to those used in the study of
Einstein-Yang-Mills-Higgs multi-monopoles~\cite{Hartmann:2001ic}, we  could confirm the
existence of gravitating generalisations of the  axially symmetric
configurations in \cite{Paturyan:2005ik}.
This includes also
 black hole solutions whose horizon has a spherical topology, but geometrically differs from a sphere.
However, due to the highly nonlinear nature of the problem,
we could not clarify  the issue of the limiting behaviour of these
gravitating axially symmetric configurations.

 %%%%%%%%%%%%%%%%%%%%%%%%%%%%%%%%%%%%%%%%%%%%%%%%%%%%
\section*{Acknowledgements}
%%%%%%%%%%%%%%%%%%%%%%%%%%%%%%%%%%%%%%%%%%%%%%%%%%%%
We are grateful to Dieter Maison for participating in the initial stage of this project.
We thank Jutta Kunz and Michael Volkov for enlightening discussions and Yves Brihaye for useful comments on a draft of this work. 
Y.S. is very grateful to Adam Winstanley
for kind hospitality at the Department of Computer Science, National
University of Ireland Maynooth, Ireland.

This work is carried out in the framework of Science Foundation Ireland (SFI) project
RFP07-330PHY.

\end{document}